\documentclass[letter]{aa} % for the letters 
\usepackage{graphicx}
\usepackage{txfonts}
\usepackage[colorlinks,citecolor=blue]{hyperref}
\usepackage{xcolor}
\usepackage[normalem]{ulem}

\begin{document}

   \title{Effect of magnetic field inclination on black hole jet power and particle acceleration}
   
 \author{Enzo Figueiredo
          \inst{1}
          \thanks{enzo.figueiredo@univ-grenoble-alpes.fr}
          \and
          Beno\^it Cerutti\inst{1}
          \and
          Kyle Parfrey \inst{2}
          }

\institute{Univ. Grenoble Alpes, CNRS, IPAG, 38000 Grenoble, France\\
           \and
           Princeton Plasma Physics Laboratory, Princeton, NJ 08540, USA
           }

\date{Received June 5th, 2025; accepted July 21st, 2025.}

  \abstract{Rotating black holes are known to launch relativistic jets and accelerate particles provided they accrete a magnetized plasma. However, it remains unclear how the global magnetic field orientation affects the jet powering efficiency. Here, we propose the first kinetic study of a collisionless plasma around a Kerr black hole that is embedded in a magnetic field inclined with respect to the black hole's spin axis. Using three-dimensional general-relativistic particle-in-cell simulations, we show that while oblique magnetic field configurations significantly reduce the jet power, particle acceleration remains highly efficient regardless. This suggests that black holes producing a weak jet could still be  bright sources of nonthermal radiation and cosmic rays.}

\maketitle
\section{Introduction}
Black holes power relativistic jets from galactic scales with stellar-mass black holes in X-ray binaries \citep{Mirabel1994,Mirabel1999,Fender2006}, to extragalactic scales with supermassive black holes in active galactic nuclei \citep{Urry1995, 2019ARA&A..57..467B}. The most promising mechanism to explain how black holes launch jets was proposed by \cite{Blandford&Znajek1977}. In this model, the jet power depends on the black hole's spin and on the amount of magnetic flux crossing its horizon. This mechanism was first studied numerically with general-relativistic magnetohydrodynamic (GRMHD) simulations \citep{2001MNRAS.326L..41K, 2002Sci...295.1688K}. In addition to the fluid-like behavior present in MHD, the near-horizon environment of a black hole may produce nonthermal radiation suggesting that particle acceleration occurs there \citep{2022ApJ...930L..16E}. This is corroborated by recent general-relativistic particle-in-cell (GRPIC) simulations that showed that efficient particle acceleration takes place within the black hole's magnetosphere via magnetic reconnection or polar-cap discharge \citep{Parfrey2019, Crinquand2020, Crinquand2021, 2021PhRvL.127e5101B, 2025ApJ...985..159Y}.

Particle acceleration and the jet powering mechanism have been studied in axisymmetric setups, where the black hole's spin direction is aligned with the surrounding magnetic field orientation \citep{Parfrey2019}. Although this was motivated for both modeling and numerical simplicity, many systems may have a preferred direction for the magnetic field that may be inclined to the black hole's spin. Wind-fed black holes may be such examples, as well as neutron star--black hole binaries, high-mass X-ray binaries, and the Gaia black holes \citep{ElBadry2023}. In the Galactic Center, Sagittarius A* (Sgr A*) may accrete stellar winds from the surrounding stars although it is still unclear whether it produces a jet \citep{Royster2019}. Isolated black holes fed by the interstellar medium \citep{Barkov2012,Kin2025} could be another favorable configuration for an inclined magnetic field. Inclined magnetospheres have been studied using GRMHD simulations \citep{Palenzuela2010, Ressler2021,Ressler2023,James2024}. These studies clearly show that jet launching is weaker when the magnetic field and the black hole's spin are inclined, but cannot capture self-consistently nonideal plasma physics and particle acceleration.

In this work, we present the first kinetic plasma study of a black hole embedded in an inclined magnetic field to understand how the jet forms and its impact on particle acceleration. We perform three-dimensional GRPIC simulations using the code \texttt{GRZeltron} \citep{cerutti2013,Parfrey2019}. We assume a Kerr spacetime, using the Kerr-Schild spherical coordinates $(t,r,\theta,\phi)$, and solve the particles and field equations in the 3+1 formalism \citep{Komissarov2004}. We use geometrized units $G = c = M = 1$, $M$ being the black hole mass; the characteristic units of length and time are $r_g = G M /c^2 $ and $t_g = r_g/c$ respectively. 
\begin{figure*}
    \centering
    \includegraphics[width=2\columnwidth]{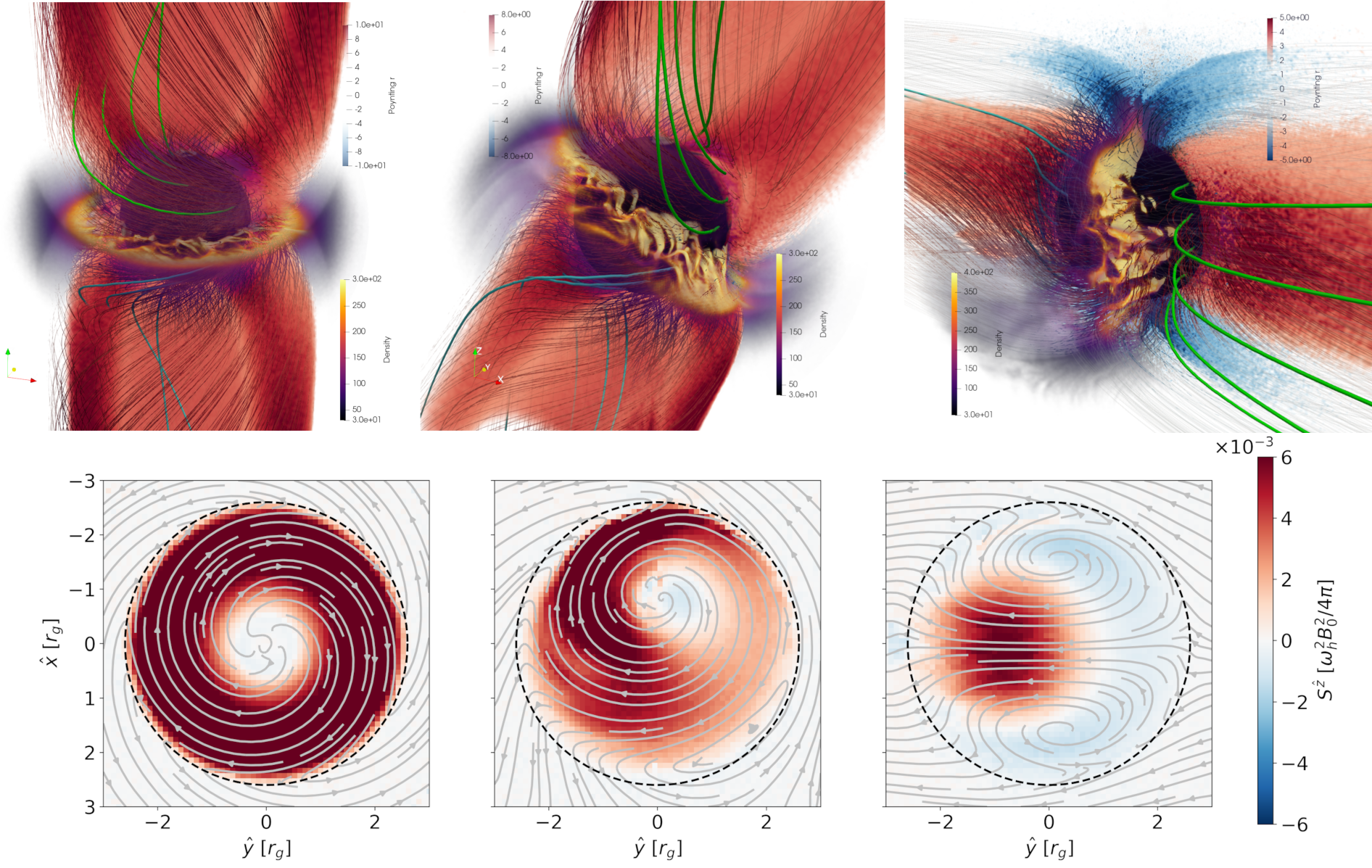}
    \caption{\emph{Top:} 3D visualizations of black hole magnetospheres with spin $a = 0.99$. We show magnetic field lines (with a few highlighted in green/blue) and volume renderings of the plasma number density over the whole domain and of the radial Poynting flux density ($S^{r}$) on half of the domain ($\phi \in [0,\pi]$). The inclination angle values are, from left to right, $\chi$ = 0°, $\chi$ = 30°, and $\chi$ = 85°. \emph{Bottom:} The outgoing Poynting flux density in a cross section of the jet ($S^{\hat{z}}$) at $r=2r_{g}$ in a plane ($O\hat{x}\hat{y}$) perpendicular to the jet axis ($\hat{z}$), for the corresponding inclination angles. Magnetic field lines in this plane are represented in gray and the dashed lines enclose the jet area.}
    \label{fig:3Drendering}
\end{figure*}

\section{Methods}

The computational domain covers $r \in [0.98 r_h,16 r_g]$, $\theta \in [0.013\pi,0.987\pi]$, $\phi \in [0,2\pi] $, where $r_h = r_g (1 + \sqrt{1-a^2})$ is the event horizon radius, and $a$ is the black-hole spin parameter. The grid consists of $N_r \times N_\theta \times N_\phi = 1024 \times 512 \times 768$ cells, equally spaced in $\log{r}$, $\theta$ and $\phi$, such that the grid is finer near the horizon.
We assume axial symmetry as $\theta$ boundary conditions and particles passing through the axis are absorbed.
Waves and particles are absorbed in a layer at the outer radial boundary of the box in which the magnetic field lines are matched to the initial field.
The initial electromagnetic configuration follows the solution by \cite{Bicak&Janis1985}. It is the electromagnetic configuration in vacuum for a rotating black hole embedded in an asymptotically uniform magnetic field $B_0$ inclined at an angle $\chi$. Similarly to the Wald solution ($\chi=0^{\rm o}$, \citealt{Wald1974}), there are regions where the electric field $\mathbf{D}$ has a component parallel to the magnetic field $\mathbf{B}$, where $\mathbf{D}$ and $\mathbf{B}$ are measured by the local fiducial observers (FIDOs, \citealt{Komissarov2004}). 

We aim to study a magnetosphere in a nearly ideal force-free regime, which implies that any electric field parallel to the magnetic field is screened and that the plasma is magnetically dominated. Thus the FIDO-measured pair number density $n$ should exceed the same observer's Goldreich-Julian number density $n_0 = \omega_h B_0 /4 \pi e$ \citep{Goldreich1969} and the plasma magnetization $\sigma = B^2/4 \pi n m \gg 1$, where $m$ and $e$ are the electron mass and charge, and $\omega_h = a/2r_h$ is the angular velocity of the event horizon.
Plasma injection is accomplished by using an ad hoc prescription. The domain in which pairs can be injected covers $r\in [1.2 r_h, 6 r_g]$. In each cell, the magnetization $\sigma$ is computed: if it is higher than a ceiling value $\sigma_c$, a pair is injected in that cell with a density $\delta n = \mathcal{R}  n_0 B^2 / B^2_0$, where $\mathcal{R}$ is the injection rate. The particles are injected with a velocity randomly drawn from a Maxwellian moving with the FIDO and having a temperature $k T = 0.5  m$. Although plasma injection is not modeled self-consistently, it is sufficient to capture most of the features of a force-free magnetosphere \citep{Parfrey2019}. We are not able, however, to reproduce the spark-gap physics observed with realistic pair injection \citep{Crinquand2020}, this matter is left to a future study. 
We achieve a force-free-like regime in our simulations by setting the dimensionless magnetic field strength $\tilde{B}_0\equiv eB_0 r_g/m = 500$, and the injection parameters $\sigma_c = 580$ and $\mathcal{R} = 1$.  The resolution of the characteristic plasma skin depth $d_0$ on the event horizon is $d_0 = \sqrt{m/4\pi n_0 e^2} \sim 30 \Delta r \sim 10 r_h \Delta \theta \sim 7 r_h \Delta \phi$.

\section{Results}

In this work, we explore four inclinations, $\chi$ = 0° (aligned), 30°, 60° and 85° and consider a high ($a=0.99$) and a moderate ($a=0.7$) spin value. An equivalent axisymmetric two-dimensional simulation is also performed for $a=0.99$. 
After a transient regime, the simulations reach a quasi-steady state at physical time between $45\, t_g$ and $80\, t_g$, depending on the inclination, from which point the global features appear static. All the results and the figures shown in this letter were processed after this steady state is reached.
In the top panel of FIG.~\ref{fig:3Drendering}, we provide 3D visualizations of the magnetosphere with a spin $a=0.99$ for different inclination values. The moderate spin simulations, with $a=0.7$, show the same qualitative behavior, although less pronounced because of the smaller size of the ergosphere. The aligned case matches qualitatively with expectations from 2D \citep{Parfrey2019}: the magnetic field lines are dragged within the ergosphere, eventually entering the horizon and assuming a helical shape.
A magnetic discontinuity appears between the upper ($\theta < \pi/2$) and the lower hemispheres ($\theta >\pi/2$) that is supported by an equatorial current layer in the ergosphere. This current layer then fragments due to the onset of the tearing instability mediating fast magnetic reconnection. 
Inclined simulations also show the field lines threading the horizon, but their geometry is altered as they asymptotically follow the magnetic field's direction at infinity. Indeed, the jet's direction is fully determined from the launching regions to infinity by the orientation of the large-scale field. A current layer still forms within the ergosphere, being transverse to the asymptotic magnetic field orientation (thus no longer in the  equatorial plane) and showing the development of magnetic reconnection. For high inclination values, we observe that the current sheet is also warped along the azimuthal direction close to the black hole's equator. 

To trace the regions of electromagnetic energy extraction within the magnetosphere, we represent the radial Poynting flux density ($S^r$) for an observer at infinity in the top panel of FIG.~\ref{fig:3Drendering}. The bottom panel displays several cross sections perpendicular to the jet axis for different magnetic field inclinations. For $\chi$ = 0°, the Poynting flux distribution is axisymmetric and concentrated in a sheath around the jet's core. The jet section shows a single magnetic cell, inherited from the field's helical shape, that supports an outgoing electric current in the jet's core (not shown here).
Although the jet still follows the magnetic field direction for inclined configurations, its internal structure is significantly altered. The jets' cross sections show that the magnetic geometry loses the helical structure for a more complex shape, in which the two hemispheres of the magnetosphere actually contribute to the same side of the jet. This is clearly visible for $\chi$ = 85°: the two counter-rotating helices of the magnetic field follow from the connection of the field lines to both hemispheres, each one sending a current with an opposite sign (see the dual magnetic cell structure in FIG.~\ref{fig:3Drendering}). The electromagnetic flux thus no longer appears to lie within a sheath but becomes less symmetric for moderate inclinations (e.g. $\chi$ = 30°), shrinking one side of the sheath while expanding the other. For $\chi$ = 85°, the maximum power appears rather concentrated within the jet core, where the transverse magnetic field strength is the highest. One can also notice the negative Poynting flux contribution coming from polar regions of the magnetosphere that accounts for electromagnetic energy being instead absorbed by the black hole.

\begin{figure}
    \centering
    \includegraphics[width=0.95\columnwidth]{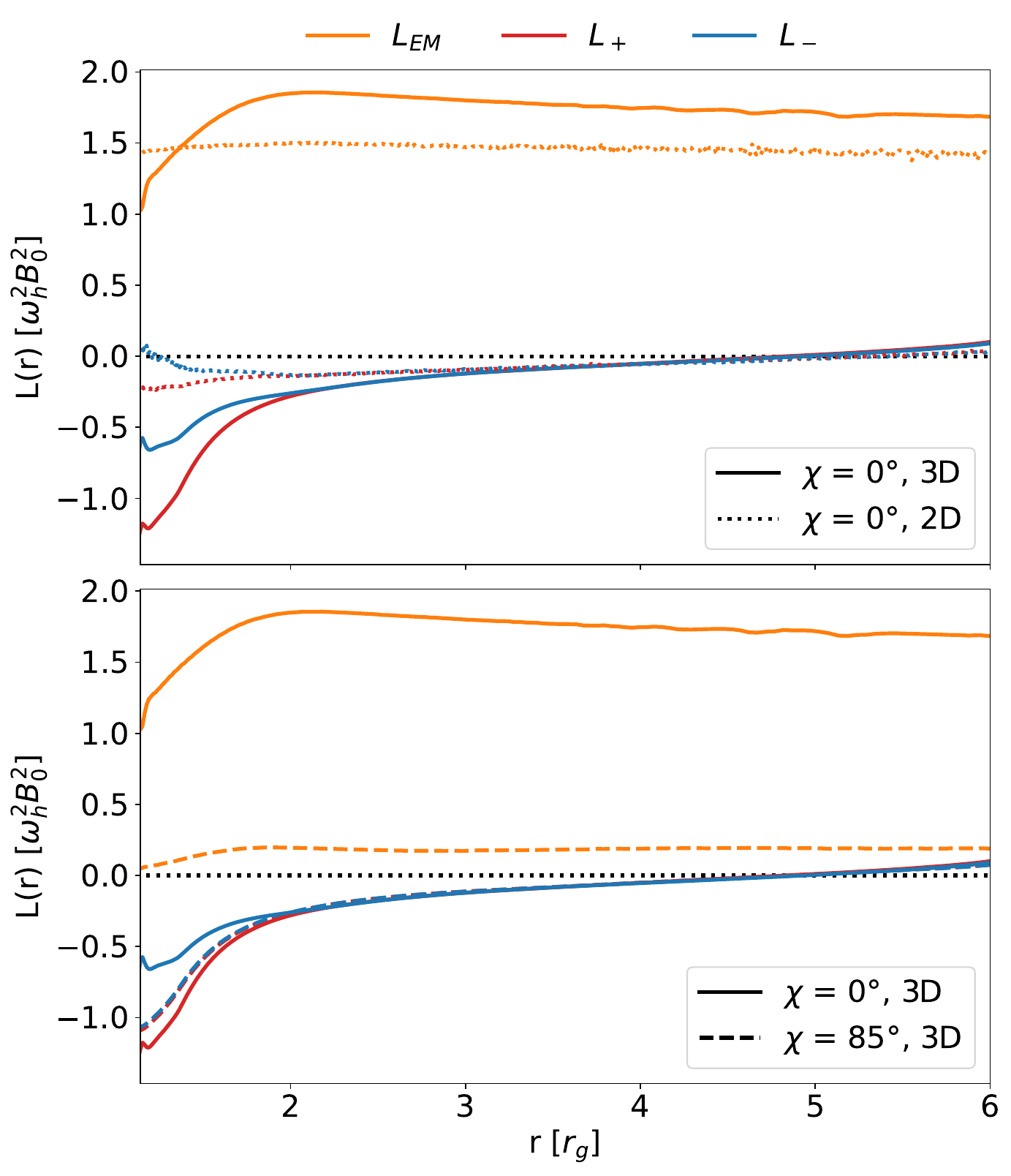}
    \caption{Radial dependence of the outgoing energy flux as measured by an observer at infinity; $a=0.99$ in all cases. The yellow lines represent the electromagnetic powers ($L_{\rm EM}$), the blue and red lines indicate respectively the electrons' and positrons' kinetic power ($L_\pm$). \emph{Top panel:} comparison between the aligned 3D simulation (solid lines) and the 2D axisymmetric simulation (dotted lines). \emph{Bottom panel:} comparison between the $\chi$ = 0° simulation (solid lines) with the inclined $\chi$ = 85° run (dashed lines). All curves were averaged over 4 $t_g$.}
    \label{fig:flux-radius}
\end{figure}

These simulations also allow us to understand the added value of 3D simulations even for an aligned magnetosphere. Here, as opposed to in 2D studies, the toroidal component of the magnetic field is not just an out-of-plane component but can also reconnect efficiently in the plasmoid-dominated regime. We clearly observe that it actually dominates over reconnection of the radial field component by a factor 4--5. This explains the large difference in the amount of dissipation and particle acceleration we observe in the top panel of FIG.~\ref{fig:flux-radius}. This figure shows the dependence of the electromagnetic and particle outgoing powers on the radial coordinate in the $\chi$ = 0° and $a=0.99$ simulation; these quantities are respectively defined as
\begin{align}
    L_{\rm EM}(r) &= \iint \sqrt{\gamma} S^r \mathrm{d} \theta \mathrm{d}\phi, \\ 
    L_{\rm \pm}(r) &= \iint \sqrt{\gamma} \left\langle e^\pm_{\infty} v^r_\pm \right\rangle n_\pm \mathrm{d} \theta \mathrm{d}\phi,
\end{align}
where the $+$ ($-$) index indicates  the positron (electron) kinetic energy flux, $\gamma$ is the determinant of the spatial 3-metric, $n_\pm$ is the number density measured by the FIDO, $v^r_\pm$ is the particle's radial coordinate-basis 3-velocity and $e^\pm_{\infty}=-u^\pm_t$ is the particle's energy as measured by an observer at infinity ($u_\mu$ being the particle's covariant 4-velocity), the product of these latter two quantities being averaged over each cell.
In 3D, a higher fraction of the total power is being carried by the particles inside the ergosphere due to magnetic dissipation, leading to a reduction in the electromagnetic power with decreasing radius. The electromagnetic power appears slightly higher asymptotically ($\sim 15\%$) than in the 2D case which may be due to a slight difference in the electromagnetic fields' geometry caused by the more vigorous magnetic reconnection occurring in the current layer.
The asymmetry in the electron and positron powers within the ergosphere is explained by the global electric polarization of the magnetosphere induced by the black hole's rotation. 
If the spin or the magnetic field direction were inverted, we would see the opposite trend for the two species. These results show that 3D models are necessary if one wishes to capture the correct level of magnetic dissipation and particle energization in the magnetosphere.

Inclining the magnetic field with respect to the black hole's spin axis also has a large effect on energy transport within the magnetosphere. In the bottom panel of FIG.~\ref{fig:flux-radius} we present the same physical quantities as discussed above, but for inclinations $\chi$ = 85° and $\chi$ = 0° and with a spin $a=0.99$. We see a significant drop in the Poynting flux when the magnetic field is inclined, mostly because the field lines are less twisted in the oblique magnetosphere.
The drop of the electromagnetic power within the ergosphere with respect to its asymptotic value is also relatively larger for the $\chi$ = 85° case (i.e., a nearly $80\%$ drop, as opposed to $\sim 40\%$ for $\chi$ = 0°) and can be explained by the contributing negative Poynting power observed around the horizon's polar region in FIG.~\ref{fig:3Drendering}. In contrast, particle energization is nearly unaffected by inclination.  Although its geometry changes with magnetic inclination, the properties of the current layer and of the upstream plasma remain mostly unchanged, leading to a similar conversion of Poynting flux to energetic particles. This is why the positron kinetic energy fluxes are very similar for $\chi$ = 0° and $\chi$ = 85°. 
The inclination also appears to quench the polarization of the magnetosphere described in the last paragraph, analogously to pulsar magnetospheres \citep{Philippov2018}, thus the same trend is observed between the electron and positron power for $\chi$ = 85°.

\begin{figure}
    \centering
    \includegraphics[width=0.9\columnwidth]{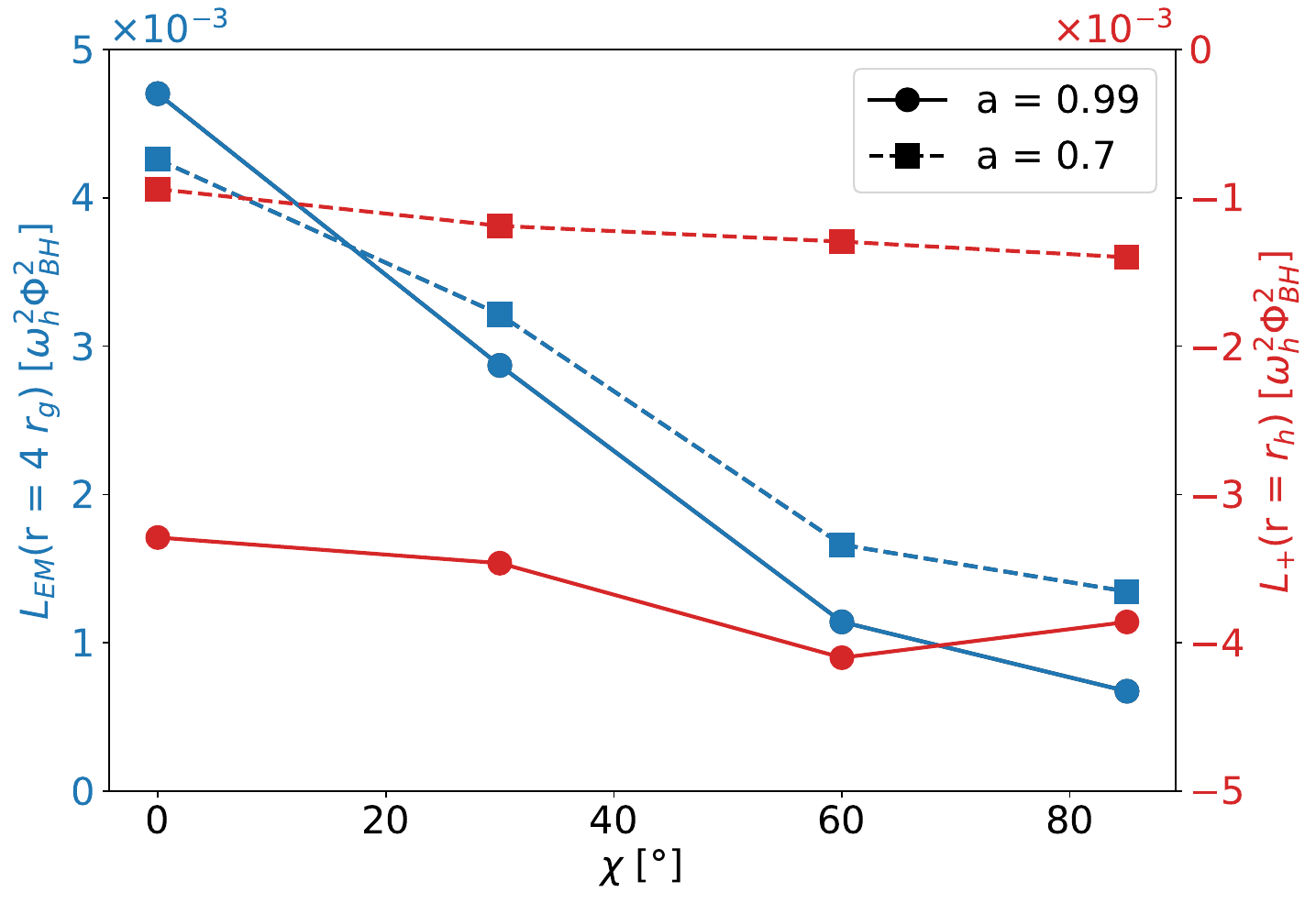}
    \caption{Energy flux dependence  with the inclination angle $\chi$ for both the electromagnetic power ($L_{\rm EM}$, blue) and the positron kinetic power ($L_+$, red). Squares represent values for a spin $a=0.7$ and circles for $a=0.99$.}
    \label{fig:flux-chi}
\end{figure}

The dependence of the magnetospheric energetics on the inclination is summed up in FIG.~\ref{fig:flux-chi}. The Poynting power, at a radius $r = 4 r_g$, is represented for all inclinations and spin values explored in this work, and it is normalized to $\omega^2_h \Phi^2_{\rm BH}$, with  $\Phi_{\rm BH} =  \iint \textrm{d}\Omega |B^r|/2$ the magnetic flux through the event horizon ($L_{\rm EM}/\omega^2_h \Phi^2_{\rm BH}$ reaches a steady-state value for all simulations). We find that the horizon magnetic flux is weakly dependent on both the spin and the inclination, lying in the range $\Phi_{\rm BH}/2\pi r^2_g B_0 \sim 2.5 - 3$ in all simulations. However, we observe that both Poynting power curves show the same trend toward a significant decrease of the electromagnetic power with the inclination angle $\chi$. The difference between the aligned and the $\chi$ = 85° cases reaches a factor $\sim 5$ for $a=0.99$ and $\sim 3$ for $a=0.7$. The significant drop in power means that the jet power in the $\chi=$ 85°, a = 0.99 simulation is comparable to that of the aligned a = 0.7 simulation. This creates a degeneracy in the jet power between a lower spin value and a highly inclined magnetic field. This is a larger effect than in force-free studies \citep{Palenzuela2010} where the electromagnetic power decreased by a factor of $\sim 2$ (at $a=0.7$). The similarity of the trends for the two spin values shows the robustness of the dependence in $\omega^2_h \Phi^2_{\rm BH}$ and in $\chi$, however one should remember that $\omega_h$ reaches a higher value for a higher spin and thus a high-spin black hole should power a stronger jet than one with a lower spin, with $\chi$ fixed.

\begin{figure}[h]
    \centering
    \includegraphics[width=\columnwidth]{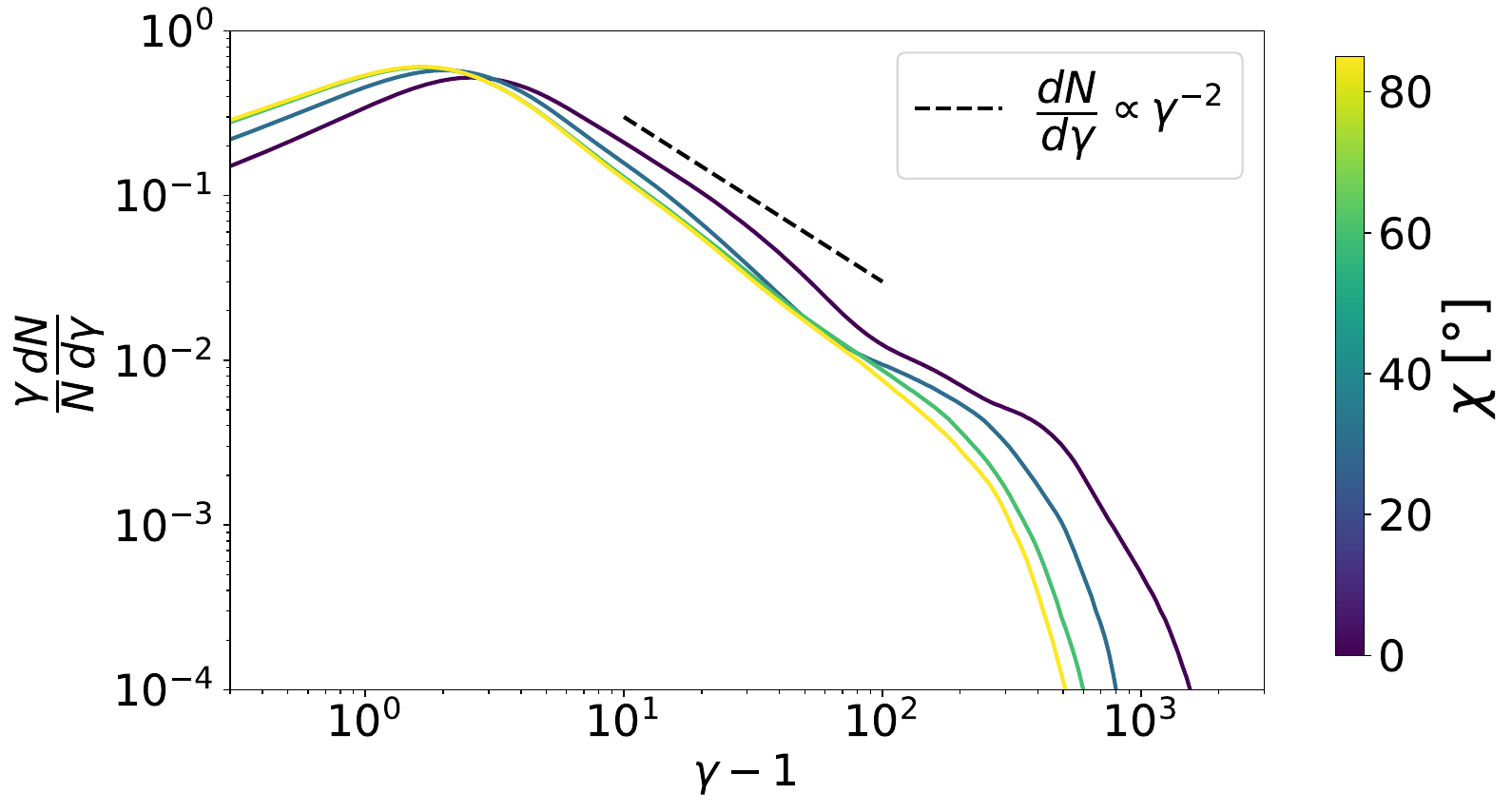}
    \caption{ Positrons' energy distributions for different $\chi$ values with $a=0.99$. Power law scaling $\mathrm{d} N/\mathrm{d} \gamma \propto \gamma^{-2}$ is shown in dashed lines.}
    \label{fig:spectra}
\end{figure}

The positron power on the black hole horizon is also represented on FIG.~\ref{fig:flux-chi}. As previously shown, a reconnecting current sheet always appear within the ergosphere and thus particles should be energized in the same way. This is why we do not observe a strong dependence in the positron power for different inclination angles. The clear higher absolute energies reached in the $a=0.99$ case compared to the $a=0.7$ case are due to the different sizes of the ergosphere: there is more room in the $a=0.99$ case to accelerated particles before they reach the event horizon. Those results demonstrate that although not being an efficient jet engine, an inclined magnetosphere should still be an efficient particle accelerator. We noticed that the particle energy distributions (see FIG.~\ref{fig:spectra}) featured nearly identical power law tails for all inclinations, with indices and cutoff energies consistent with local studies of magnetic reconnection \citep{2014ApJ...783L..21S, 2016ApJ...816L...8W}.

\section{Conclusions}

This work allows us to understand several key aspects of black hole magnetospheres. Consistently with previous studies, we show that an inclined magnetic configuration with respect to the black hole spin  significantly decreases the jet power. However, kinetic simulations are crucial to show that this weakening of the jet is not associated with a decrease in particle energization within the magnetosphere. Instead, magnetic reconnection appears to be very robust at providing nonthermal particle acceleration. In that sense, black holes not powering a jet because of a suboptimal magnetic configuration could still leave high-energy radiative signatures; this may be relevant to the case of Sgr A*. The different magnetic geometry induced within the jet is non-trivial and should leave imprints in the polarimetric signature.
A byproduct of our study is a demonstration that 3D simulations of black hole magnetospheres are essential to realistically model the dissipation within the ergosphere. In 3D, significantly more magnetic energy is released into particle acceleration, which has implications for the magnetosphere's radiative flux estimation.

Our work is however limited by the nonphysical pair creation we used. A complementary study would present more realistic pair creation using self-consistent Monte Carlo modeling of high-energy photons and inverse-Compton scattering \citep{Crinquand2020, 2025ApJ...985..159Y} and would allow us to understand how the spark-gap physics is altered by this non-axisymmetric setup. The shape of the current layer, the expected main source of high-energy radiation, would lead to different light curves or polarization signatures that should be valuable constraints on the black hole's surrounding environment. A setup with a time-varying magnetic field direction would also be relevant to the case of a black hole embedded in a pulsar wind.

\begin{acknowledgements}
This project has received funding from the European Research Council (ERC) under the European Union’s Horizon 2020 research and innovation program (Grant Agreement No. 863412). Computing resources were provided by TGCC under the allocations SS010415385 and A0170407669 made by GENCI. KP acknowledges support from the Laboratory Directed Research and Development Program at Princeton Plasma Physics Laboratory, a national laboratory operated by Princeton University for the U.S.\ Department of Energy under Prime Contract No.\ DE-AC02-09CH11466. We are thankful to the referee for their comments that helped us clarifying some of our results.
\end{acknowledgements}

\bibliographystyle{aa}
\bibliography{bibliography}

\end{document}